\documentclass[aps,prd,twocolumn,showpacs,superscriptaddress,10pt]{revtex4-1}

\usepackage{acronym}
\usepackage{amsfonts,amsmath,amssymb}
\usepackage{bm}
\usepackage{cases}
\usepackage{comment}
\usepackage[T1]{fontenc} 
\usepackage{graphicx}
\usepackage{indentfirst}
\usepackage{mathrsfs}
\usepackage{multirow}
\usepackage{scrextend}
\usepackage{hyperref}
\usepackage{xcolor}
\usepackage{soul}
\usepackage[most]{tcolorbox}
\usepackage{float}
\usepackage{ulem}

\begin{document}

\title{Quasinormal modes of Reissner-Nordstr\"om-AdS black holes under physical field-vanishing boundary conditions}

\author{Hui-Fa Liu}
\email{hfaliu@163.com}
\affiliation{Beijing University of Technology, Beijing 100124, China}

\author{Qi Su}
\email{sqphysics@outlook.com}
\affiliation{Beijing University of Technology, Beijing 100124, China}

\author{Ding-fang Zeng}
\email{dfzeng@bjut.edu.cn}
\affiliation{Beijing University of Technology, Beijing 100124, China}

\date{\today}

\begin{abstract}

Boundary conditions play a key role in determining the perturbation behavior of a black hole. Motivated by two guiding principles for single-field perturbations --- the non-deformation of the boundary metric and the vanishing of electromagnetic energy flux at the AdS boundary --- we impose a boundary condition for Reissner-Nordström-AdS (RN-AdS) black holes requiring both the metric and electromagnetic field-strength perturbations to vanish at the AdS boundary, which we term the physical field-vanishing (PFV) condition. Using the formulas for perturbation reconstruction, we translate the PFV condition into boundary conditions on the master functions: Dirichlet-type for odd-parity modes and Robin-type for even-parity modes. With these boundary conditions, we compute the quasinormal frequencies of RN-AdS black holes and identify new spectral features. The PFV prescription introduced here could be applied to other multifield perturbation systems in asymptotically AdS spacetimes.

\end{abstract}

\maketitle

\section{Introduction}

Perturbations of a black hole give rise to a series of damped oscillatory modes known as quasinormal modes (QNMs)~\cite{Chandrasekhar:1983,Nollert:1999ji,Kokkotas:1999bd,Berti:2009kk,Konoplya:2011qq,Bolokhov:2025uxz,Berti:2025hly,Bourg:2024jme,Bourg:2025lpd}, which are characterized by its key parameters. According to black hole perturbation theory, the essential features of the perturbations are encoded in the master equations. Solving these equations to obtain the QNM spectrum requires physically reasonable boundary conditions on the master functions. Typically, one imposes purely ingoing conditions at the event horizon and a proper condition at spatial infinity (or the cosmological horizon) that is determined by the asymptotic structure of the spacetime. 

The perturbations of black holes in the asymptotically AdS spacetime attract much attention, motivated by the AdS/CFT duality. By this duality, the damping of bulk perturbations corresponds to the relaxation of thermal fluctuations in the boundary conformal field theory (CFT). The confining nature of AdS spacetime implies that the boundary at infinity acts as a perfect reflector, and the boundary conditions there must therefore reflect this asymptotic feature. A common choice, motivated partly by simplicity, is to impose Dirichlet boundary conditions on the master functions~\cite{Horowitz:1999jd,Cardoso:2001bb,Cardoso:2003cj,Konoplya:2002ky,Wang:2000gsa,Berti:2003ud}. However, imposing boundary conditions directly on the physical fields --- by which we refer to the metric and electromagnetic field perturbations that determine boundary observables --- rather than on the master functions, is particularly natural in the holographic context~\cite{Cardoso:2013pza}.

Two guiding principles for single-field perturbations were proposed in previous works. For the gravitational perturbations, Michalogiorgakis and Pufu~\cite{Michalogiorgakis:2006jc} proposed the non-deformation of the boundary metric for scalar-sector perturbations in the Kodama-Ishibashi formalism~\cite{Kodama:2000fa,Kodama:2003jz,Ishibashi:2003ap,Kodama:2003kk}. This principle results in Robin-type boundary conditions for the master functions and has been further extended to rotating AdS black holes~\cite{Dias:2013sdc}. For the Maxwell perturbations, Wang \textit{et al.}~\cite{Wang:2015goa,Wang:2015fgp,Wang:2021uix} advocated the requirement of vanishing energy flux at the AdS boundary, leading to two Robin conditions for the Teukolsky master functions, one of which reproduces the standard spectrum while the other uncovers a new family of modes. Later analyses in the Regge-Wheeler-Zerilli (RWZ) formalism confirmed this result~\cite{Wang:2021upj}: although Maxwell perturbations for both odd and even parity obey identical master equations, the vanishing-flux principle leads to different boundary conditions on the master functions, and therefore to different spectral behavior.

In this paper, we extend the study to the linear perturbations of an RN-AdS black hole in the RWZ formalism. In this case, the master functions are constructed as linear combinations of the metric and electromagnetic perturbations as well as their derivatives~\cite{Liu:2025reu}. As mentioned above, the guiding principles for boundary conditions require that for purely gravitational perturbations, $h_{\mu\nu}\to0$ at infinity, while for pure Maxwell perturbations, the condition is the vanishing of the energy flux at the boundary. However, extending these principles to the coupled system is nontrivial. Directly imposing a vanishing-flux condition for the gravitational sector would involve the second-order effective energy-momentum tensor, which depends on intricate quadratic combinations of first-order perturbations~\cite{Cardoso:2013pza}. This complexity motivates a simpler alternative: applying a non-deformation condition that simultaneously requires both the metric and electromagnetic field-strength perturbations to vanish at the AdS boundary. This approach not only avoids higher-order complications but also, as we will show, ensures that the electromagnetic energy flux vanishes at infinity. We therefore introduce this unified boundary condition for the coupled system, termed the PFV condition, which naturally generalizes the non-deformation principle to multiple fields.

To implement this prescription, we reconstruct the perturbations in the RWZ gauge~\cite{Regge:1957td,Zerilli:1970se,Zerilli:1970wzz,Zerilli:1974ai}, expressing the metric and electromagnetic potential perturbations in terms of the master functions and their derivatives~\cite{Gleiser:1995gx,Gleiser:1998rw,Jhingan:2002kb,Nakano:2007cj,Lenzi:2024tgk}. This reconstruction makes it possible to translate the PFV conditions into explicit boundary conditions on the master functions. It is also indispensable for future second-order perturbation analyses, where quadratic combinations of the first-order perturbations act as sources in the master equations~\cite{Loutrel:2020wbw,Spiers:2023cip,Spiers:2023mor,Brizuela:2006ne}. Finally, using these explicit boundary conditions, we compute the QNMs of RN-AdS black holes and identify novel spectral features that arise from the coupled dynamics. 

The remainder of this paper is organized as follows. Section \ref{Linear-perturbations-of-RN-AdS-black-holes} reviews the RN-AdS black hole background, the harmonic decomposition of perturbations, and the master equations. Section \ref{Reconstruction-of-the-perturbations} presents the reconstruction of perturbations, while Section \ref{Boundary-conditions-on-master-functions} derives the mapping from the PFV condition to boundary conditions on the master functions. In Sec. \ref{Numerical-method-and-results}, we perform the numerical computation of quasinormal frequencies and analyze the resulting spectral properties. Section \ref{Discussion} concludes with a summary and discussion. Throughout the paper, we adopt natural units and the metric signature $(-,+,+,+)$.


\section{Linear perturbations of RN-AdS black holes}
\label{Linear-perturbations-of-RN-AdS-black-holes} 

\subsection{Background}

We begin with the four-dimensional RN-AdS black hole background. The spacetime metric $g_{\mu\nu}$ and the electromagnetic potential $A_{\mu}$ are given by
\begin{align}
&g_{\mu\nu}dx^{\mu}dx^{\nu}=-f(r)dt^2+\frac{1}{f(r)}dr^2+r^2d\Omega^2\,,
\label{gbar}
\\
&f(r)=1-\frac{2M}{r}+\frac{Q^2}{r^2}+\frac{r^2}{L^2}\,,\quad A_{\mu}dx^{\mu}=-\frac{Q}{r}\,dt \,,
\label{Abar}
\end{align}
where $d\Omega^2=d\theta^{2}+\sin^{2}\theta d\phi^{2}$. Here, $M$ and $Q$ represent the mass and charge of the black hole, respectively, and $L$ is the AdS curvature radius. The cosmological constant is given by $\Lambda=-3/L^2$. The event horizon radius $r_+$ is defined by $f(r_+)=0$. It is then convenient to rewrite the mass as
\begin{align}
M=\frac{1}{2}\Big(r_{+}+\frac{Q^2}{r_+}+\frac{r_{+}^3}{L^2}\Big)\,.
\end{align}
The Hawking temperature is given by
\begin{align}
T=\frac{f'(r_+)}{4\pi}=\frac{1}{4\pi r_+}\Big(1-\frac{Q^2}{r_{+}^2}+\frac{3r_{+}^2}{L^2}\Big)\,.
\label{temp}
\end{align}
The extremal configuration corresponds to $T=0$, where the charge reaches its extremal value
\begin{align}
Q^2_{\text{ext}}=r_{+}^2\Big(1+\frac{3r_{+}^2}{L^2}\Big)\,.
\label{Qext}
\end{align}

\subsection{Decomposition of perturbations}

We now consider small perturbations on the RN-AdS background. The full spacetime metric and electromagnetic potential are expanded to linear order as
\begin{align}
\tilde{g}_{\mu\nu}=g_{\mu\nu}+h_{\mu\nu}\,,\quad \tilde{A}_{\mu}=A_{\mu}+a_{\mu}\,,
\label{pert}
\end{align}
where $h_{\mu\nu}$ and $a_{\mu}$ denote the metric and electromagnetic potential perturbations, respectively. The perturbation of the electromagnetic field strength is
\begin{align}
\mathfrak{f}_{\mu\nu} =\partial_{\mu}a_{\nu}-\partial_{\nu}a_{\mu}\,. \label{fdef}
\end{align}
The fields $h_{\mu\nu}$ and $\mathfrak{f}_{\mu\nu}$ are coupled through the linearized Einstein--Maxwell equations, forming a dynamical system that mixes the gravitational and electromagnetic degrees of freedom.

Exploiting the spherical symmetry of the background, all perturbations are decomposed into tensor spherical harmonics, which are constructed from the scalar spherical harmonics $Y_{\ell m}(\theta,\phi)$. This separates the system into odd-parity (axial) and even-parity (polar) sectors. For a given harmonic index $\ell$ and in the RWZ gauge, the ans\"atz for the metric and electromagnetic potential perturbations take the form
\begin{widetext}
\begin{align}
h^{\rm odd}_{\mu\nu}&=
\begin{pmatrix}
	0 & 0 & -{h_0}(t,r)\frac{1}{\sin\theta}\frac{\partial}{\partial\phi}Y_{\ell m} &  {h_0}(t,r)\sin\theta\frac{\partial}{\partial\theta}Y_{\ell m} \\
	0 & 0 & -{h_1}(t,r)\frac{1}{\sin\theta}\frac{\partial}{\partial\phi}Y_{\ell m} & {h_1}(t,r)\sin\theta\frac{\partial}{\partial\theta}Y_{\ell m} \\
	Sym & Sym & 0 & 0 \\
	Sym & Sym & 0 & 0 \\
\end{pmatrix}\,,
\label{oddpergen}
\\
a^{\rm odd}_{\mu}&=(0\,,\,0\,,\,-B(t,r)\frac{1}{\sin\theta}\frac{\partial}{\partial\phi}Y_{\ell m}\,,\,B(t,r)\sin\theta \frac{\partial}{\partial\theta}Y_{\ell m})\,,
\end{align}
and
\begin{align}
h^{\rm even}_{\mu\nu}&=
\begin{pmatrix}
	f(r){H_0}(t,r)Y_{\ell m} & {H_1}(t,r)Y_{\ell m} & 0 & 0 \\
	Sym & \frac{{H_2}(t,r)}{f(r)}Y_{\ell m} & 0 & 0 \\
	0 & 0 & r^2K(t,r)Y_{\ell m} & 0 \\
	0 & 0 & 0 & r^2K(t,r)\sin^2\theta Y_{\ell m} \\
\end{pmatrix}\,,
\label{evenpergen}
\\
a^{\rm even}_{\mu}&=(E_{0}(t,r){Y}_{\ell m}\,,\,E_{1}(t,r){Y}_{\ell m}\,,\,0\,,\,0)\,.
\label{evengena}
\end{align}
\end{widetext}
Based on these ans\"atz, the linearized Einstein--Maxwell equations reduce to a set of coupled partial differential equations for the functions $h_0$, $h_1$, $B$, $H_0$, $H_1$, $H_2$, $K$, $E_0$, and $E_1$.

\subsection{Master function and equations}

The reduced equations can be diagonalized into two independent dynamical modes for each parity sector, represented by a pair of master functions $\Psi_i(t,r)$. The index $i=0,1$ labels the two modes associated with the two physical degrees of freedom of the coupled Einstein--Maxwell system~\cite{Zerilli:1974ai,Moncrief:1974gw,Moncrief:1974ng}. In this work, we adopt the explicit form of the master functions given in Ref.~\cite{Liu:2025reu}, defining these functions as suitable linear combinations of the metric and electromagnetic perturbations.

The master equations take the Schr\"odinger-like form
\begin{align}
\left[\frac{\partial^2}{\partial r_*^2}-\frac{\partial^2}{\partial t^2}-V^{\rm odd/even}_i\right] 
\Psi^{\rm odd/even}_i=0\,,\quad i=0,1\,,
\label{master-equations}
\end{align}
where the tortoise coordinate $r_*$ is defined via $dr_*/dr=1/f(r)$, and $V^{\rm odd/even}_i(r)$ are the effective potentials. For the odd-parity sector, the effective potentials and the master functions are given by
\begin{align}
V^{\rm odd}_i(r)=&f(r)\Big(\frac{\mu^2+2}{r^2}+\frac{p_{i}-6 M}{r^3}+\frac{4 Q^2}{r^4}\Big)\,,
\label{oddv}
\\
\Psi^{\rm odd}_i(t,r)=&h_0(t,r)+\frac{r}{2}\Big[\dot{h}_1(t,r)-h_0'(t,r)\Big]
\nonumber
\\
&+\Big(\frac{2Q}{r}-\frac{p_i}{2Q}\Big)B(t,r)\,,\quad i=0,1\,.
\label{psiodd}
\end{align}
We adopt simplified notations $y' \equiv \partial y(t,r)/\partial r$ and $\dot{y} \equiv \partial y(t,r)/\partial t$. For the even-parity sector, the effective potentials and the master functions are given by
\begin{align}
V^{\rm even}_i(r)=&f(r)\Big[U(r)\!+\!\Big(\frac{\mu^2r+p_i}{r}\!-\!\lambda(r)\Big)W(r)\Big]\,,
\label{evenv}
\\
\Psi^{\rm even}_i(t,r)=&\frac{r}{2} K(t,r)\!+\!\frac{\mu^2 r\!+\!p_i}{\mu^2 \lambda (r)}f(r)\left[H_0(t,r)\!-\!rK'(t,r)\right]
\nonumber
\\
&\hspace{-1.2em}-\!\frac{p_ir^2}{2\mu^2Q}\left[E_0'(t,r)\!-\!\dot{E}_1(t,r)\right]\,,\quad i=0,1\,.
\label{psieven}
\end{align}
The parameters entering the above expressions are defined as follows
\begin{align}
\mu^2&=\ell(\ell+1)-2\,,
\\
p_0&=3M-\sqrt{9M^2+4\mu^2Q^2}\,,
\\
p_1&=3M+\sqrt{9M^2+4\mu^2Q^2}\,.
\label{p-parameter}
\end{align}
In addition, the auxiliary functions $\lambda(r)$, $U(r)$, and $W(r)$ are defined by
\begin{align}
\lambda(r)&=\mu^2+ \frac{3r^2}{L^2}+3-3f(r)-\frac{Q^2}{r^2}\,,
\\
U(r)&=\frac{\mu^2+2}{r^2}+\frac{8Q^2f(r)}{r^4\lambda(r)}\,,
\label{evenfunU}
\\
W(r)&=\frac{2(\mu^2\!+\!2\!-\!f(r))}{r^2\lambda(r)}\!+\!\frac{2f(r)}{r^2\lambda(r)^2}\left(\mu^2\!+\!\frac{4Q^2}{r^2}\right)\!-\!\frac{1}{r^2}\,.
\label{evenfunW}
\end{align}
A subtlety arises in the neutral limit $Q\to0$. In this case, the master functions for $i=1$ in Eqs.~\eqref{psiodd} and \eqref{psieven} contain terms proportional to $p_{1}/Q$ and hence diverge. As discussed in Ref.~\cite{Liu:2025reu}, this divergence does not indicate a pathology but rather the need for a proper rescaling, which isolates the finite-energy Maxwell modes in the uncharged background. 

The master equations \eqref{master-equations} form the basis for implementing the physical boundary conditions and for extracting the QNM spectrum. The master functions for a single independent dynamical mode are not unique, as they can be defined in several equivalent forms related by linear transformations, similar to vacuum gravitational cases~\cite{Lenzi:2021wpc,Lenzi:2021njy}. We adopt special choices in Eqs.~\eqref{psiodd} and \eqref{psieven} so that the reconstruction in the next section can be achieved without explicit time integrals.

\section{Reconstruction of physical field perturbations}
\label{Reconstruction-of-the-perturbations}


Translating the PFV conditions into boundary conditions for the master functions is not straightforward. Independently assigning asymptotic falloffs to all perturbation components could lead to inconsistencies, since components of the metric and electromagnetic perturbations approach zero or even diverge, with different powers of $r^{-1}$. These powers must conspire to satisfy the linearized Einstein--Maxwell equations. An efficient and elegant approach is to use the reconstruction formulas, where all perturbations are expressed as linear combinations of the master functions and their derivatives. In this way, the constraints on the falloff powers of different perturbation components are naturally implemented, leaving only a minimal set of free coefficients in the master-function expansions. The PFV conditions then reduce to selecting the allowed asymptotic branch of the master functions, which uniquely specifies the resulting boundary condition. In this section, we focus on the explicit reconstruction for both parity sectors.

For odd-parity modes, combining the two master functions for $i=0,1$ defined in Eq.~\eqref{psiodd} allows us to solve for $B(t,r)$ as
\begin{align}
B(t,r)&=-\frac{2Q}{p_0-p_1}\Big[\Psi_0(t,r)-\Psi_1(t,r)\Big]\,.
\label{reconstB}
\end{align}
Here and throughout this section, the odd/even labels are suppressed for notational simplicity; no ambiguity arises. With $B(t,r)$ explicitly determined, the metric perturbations can be reconstructed as
\begin{align}
h_1(t,r)=&\frac{2}{\mu^2(p_0-p_1)}\Big[\frac{rp_1}{f(r)}\dot{\Psi}_0(t,r)-\frac{rp_0}{f(r)}\dot{\Psi}_1(t,r)\Big]\,,
\\
h_0(t,r)=&\frac{2}{\mu^2(p_0-p_1)}\Big[p_1f(r)\Psi_0(t,r)-p_0f(r)\Psi_1(t,r)
\nonumber
\\
&+p_1rf(r)\Psi_0'(t,r)-p_0rf(r)\Psi_1'(t,r)\Big]\,.
\label{reconsth}
\end{align}

For even-parity modes, suitable linear combinations of the two master functions for $i=0,1$ in Eq.~\eqref{psieven}, together with the linearized field equations, yield relations
\begin{widetext}
\begin{align}
E_1(t,r)=&-\frac{2Q}{(\mu^2+2)(p_0-p_1)}\Big[\frac{\mu^2r+p_1}{rf(r)}\dot{\Psi}_0(t,r)-\frac{\mu^2r+p_0}{rf(r)}\dot{\Psi}_1(t,r)\Big]\,,
\label{reconstE1}
\\
{E}_0(t,r)=&\frac{2Q}{(\mu^2+2)(p_0-p_1)}\Big[\frac{p_1f(r)}{r^2}\Psi_0(t,r)-\frac{p_0f(r)}{r^2}\Psi_1(t,r)-\frac{\mu^2r+p_1}{r}f(r)\Psi'_0(t,r)+\frac{\mu^2r+p_0}{r}f(r)\Psi'_1(t,r)\Big]\,,
\\
{K}(t,r)=&\frac{2}{(\mu^2+2)(p_0-p_1)}\Big\{\Big[\frac{r ({p_1}+\mu^2r)}{f(r)}{V_0}(r)-\mu^2(\mu^2+2)-\frac{{p_1} \lambda(r)}{r}\Big]\Psi_0(t,r)-2{p_1}f(r)\Psi'_0(t,r)
\nonumber
\\
&-\Big[\frac{r ({p_0}+\mu^2r)}{f(r)}{V_1}(r)-\mu^2(\mu^2+2)-\frac{{p_0} \lambda(r)}{r}\Big]\Psi_1(t,r)+2{p_0}f(r)\Psi'_1(t,r)\Big\}\,,
\\
H_1(t,r)=&\frac{2}{(\mu^2+2)(p_0-p_1)}\Big\{\Big[\frac{r^2({p_1}+\mu^2r)}{f(r)^2}{V_0}(r)-\frac{\mu^2(\mu^2+2)r}{f(r)}\Big]\dot{\Psi}_0(t,r)-2{p_1}r\dot{\Psi}'_0(t,r)
\nonumber
\\
&-\Big[\frac{r^2({p_0}+\mu^2r)}{f(r)^2}{V_1}(r)-\frac{\mu^2(\mu^2+2)r}{f(r)}\Big]\dot{\Psi}_1(t,r)+2{p_0}r\dot{\Psi}'_1(t,r)\Big\}\,.
\label{reconstH1}
\end{align}
\end{widetext}
In addition, $H_0(t,r)$ and $H_2(t,r)$ can be expressed implicitly as
\begin{align}
H_0(t,r)=&\frac{r^2}{2}K''(t,r)-\frac{r^2}{2f(r)^2}\ddot{K}(t,r)
\nonumber
\\
&\hspace{-1em}+\Big(r+\frac{r^2f'(r)}{2f(r)}\Big)K'(t,r)-\frac{\mu^2}{2f(r)}K(t,r)
\nonumber
\\
&\hspace{-1em}+\frac{2Q}{f(r)}\Big[E'_0(t,r)+\frac{2}{r}E_0(t,r)-\dot{E}_1(t,r)\Big]\,,
\label{reconstH0}
\\
H_2(t,r)=&H_0(t,r)\,.
\label{reconstH2}
\end{align}
Substituting expressions \eqref{reconstE1}--\eqref{reconstH1} into \eqref{reconstH0}--\eqref{reconstH2}, $H_0(t,r)$ and $H_2(t,r)$ can also be written explicitly in terms of the master functions $\Psi_0(t,r)$ and $\Psi_1(t,r)$.

Thus, within each parity sector, all components of the metric and electromagnetic potential perturbations can be expressed as linear combinations of the master functions and their derivatives. We expand each master function in a modal decomposition,
\begin{align}
\Psi_i(t,r)=\sum_{n}\mathrm{e}^{-\mathrm{i}\,\omega_{n,i}t}\,\Psi_{n,i}(r)\,, \quad i=0,1\,,
\label{modal-expansion}
\end{align}
where $n=0,1,2,\dots$ labels the overtones. Each label $i=0,1$ corresponds to a distinct branch of master functions, each with its own quasinormal spectrum. The perturbation fields are therefore reconstructed as linear combinations of contributions from both branches. Specifically, the reconstructions take the form
\begin{align}
h_{\mu\nu}(t,r)&= \sum_{i=0}^{1}\sum_{n}\mathrm{e}^{-\mathrm{i}\,\omega_{n,i}t}\; 
\mathcal{D}_{\mu\nu,i}\!\big[\Psi_{n,i}(r)\big], 
\label{h-recons}\\
a_{\mu}(t,r)&= \sum_{i=0}^{1}\sum_{n}\mathrm{e}^{-\mathrm{i}\,\omega_{n,i}t}\; 
\varepsilon_{\mu,i}\!\big[\Psi_{n,i}(r)\big],
\label{a-recons}
\end{align}
where $\mathcal{D}_{\mu\nu,i}$ and $\varepsilon_{\mu,i}$ are branch- and parity-dependent linear differential operators that map a radial master eigenfunction $\Psi_{n,i}(r)$ to the corresponding perturbation component. Eqs.~\eqref{modal-expansion}--\eqref{a-recons} make clear that even if each $\Psi_{n,i}(r)$ is a single-frequency QNM, the reconstructed fields generally contain multiple frequencies because they receive contributions from both branches and from multiple overtones; this leads to multi-frequency interference in the reconstructed perturbations.

\section{Boundary conditions on master functions}
\label{Boundary-conditions-on-master-functions}

In this section, we specify the boundary conditions for the master functions governed by Eq.~\eqref{master-equations}. At the AdS boundary, the PFV condition determines the boundary behavior of the master functions, while at the event horizon we impose the ingoing wave condition. 

\subsection{AdS boundary}

At the AdS boundary, the PFV condition demands that both the metric and electromagnetic field-strength perturbations vanish asymptotically, i.e.,
\begin{align}
h_{\mu\nu} \to 0\,,\quad \mathfrak{f}_{\mu\nu} \to 0 \,,\qquad r \to \infty\,.
\label{fhvanish}
\end{align}
Asymptotic analysis shows that these physical fields approach zero according to power laws. By the reconstruction idea, they can be expressed as power-law expansions of the master functions,
\begin{align}
\Psi^{\rm odd}_i(t,r)=&\mathrm{e}^{-\mathrm{i}\,\omega_i t}\Big(\alpha^{(0)}_i+\tfrac{\alpha^{(1)}_i}{r}+\tfrac{\alpha^{(2)}_i(\alpha^{(0)}_i)}{r^2}
\nonumber
\\
&+\tfrac{\alpha^{(3)}_i(\alpha^{(0)}_i,\alpha^{(1)}_i)}{r^3}+\dots\Big)\,,\quad i=0,1\,, 
\label{PsiOddSeries}
\\[0.5em]
\Psi^{\rm even}_i(t,r)=&\mathrm{e}^{-\mathrm{i}\,\omega_i t}\Big(\beta^{(0)}_i+\tfrac{\beta^{(1)}_i}{r}+\tfrac{\beta^{(2)}_i(\beta^{(0)}_i)}{r^2}
\nonumber
\\
&+\tfrac{\beta^{(3)}_i(\beta^{(0)}_i,\beta^{(1)}_i)}{r^3}+\dots\Big)\,,\quad i=0,1\,.
\label{PsiEvenSeries}
\end{align}  
The subleading coefficients, such as $\alpha^{(2)}_i$ and $\alpha^{(3)}_i$, are non-independent; they are determined recursively by the master equations and depend only on the two free parameters $\alpha^{(0)}_i$ and $\alpha^{(1)}_i$. An analogous recursive structure applies to the even-parity expansion. When we substitute these expansions into \eqref{reconstB}--\eqref{reconstH2}, the PFV condition \eqref{fhvanish} imposes algebraic constraints on the expansion coefficients. Note that in addition to the decoupling of the odd and even parity modes, the PFV condition independently constrains the combinations of the master functions and their derivatives for each branch $i=0,1$. However, in the RWZ gauge, the resulting metric and electromagnetic perturbations do not generally satisfy Eq.~\eqref{fhvanish}; specifically, some components of $h_{\mu\nu}$ or $\mathfrak{f}_{\mu\nu}$ approach constants or even grow as powers of $r$ at infinity. This arises because the theory possesses additional gauge freedom. In a general perturbative gauge, the metric reconstruction involves not only the master functions but also explicit gauge functions, as discussed in Ref.~\cite{Lenzi:2024tgk}. Choosing a specific gauge --- such as the RWZ gauge adopted here --- corresponds to fixing these functions, which vanish in the RWZ gauge. The PFV condition does not hold in the RWZ gauge, since the gauge functions vanish; however the additional gauge freedom allows us to set the perturbations vanish at the AdS boundary. Starting from the RWZ gauge, we therefore perform such an additional gauge transformation so that Eq.~\eqref{fhvanish} is implemented.

The linear perturbation admits infinitesimal diffeomorphisms and $U(1)$ gauge transformations~\cite{Bruni:1996im},
\begin{align}
h^{\rm new}_{\mu\nu}&= h_{\mu\nu}-\nabla_{\mu}\xi_{\nu}-\nabla_{\nu}\xi_{\mu}\,,\label{transformh}
\\
a^{\rm new}_{\mu}&= a_{\mu}-\xi^{\nu}\nabla_{\nu}A_{\mu}-A_{\nu}\nabla_{\mu}\xi^{\nu}+\nabla_{\mu}\eta\,,
\label{transforma}
\end{align}
where $\xi^\mu$ is the diffeomorphism generator and $\eta$ the $U(1)$ gauge parameter. For odd-parity modes, there is one independent diffeomorphism function $\Lambda_2(t,r)$, and the corresponding gauge vector takes the form
\begin{align}
\xi^{\mu}_{\rm odd} = \Big\{0,0,-\tfrac{\Lambda_2}{\sin\theta}\tfrac{\partial}{\partial\phi}Y_{\ell m}, \tfrac{\Lambda_2}{\sin\theta}\tfrac{\partial}{\partial\theta}Y_{\ell m}\Big\}^{\scriptscriptstyle T}\, .
\end{align}
For even-parity modes, there are three diffeomorphism functions $M_0(t,r), M_1(t,r), M_2(t,r)$ and one electromagnetic gauge function $\eta(t,r)$. The corresponding gauge vector and electromagnetic function are given by
\begin{align}
\xi^{\mu}_{\rm even}{=}\Big\{M_0Y_{\ell m}, M_1Y_{\ell m}, M_2\tfrac{\partial}{\partial\theta}Y_{\ell m}, \tfrac{M_2}{\sin^2\theta}\tfrac{\partial}{\partial\phi}Y_{\ell m}\Big\}^{\scriptscriptstyle\!T}, 
\\
\eta_{\rm even}{=}\eta(t,r)Y_{\ell m}(\theta,\phi)\, .
\rule{40mm}{0pt}
\end{align}
Since the field strength $\mathfrak{f}_{\mu\nu}$ is invariant under $U(1)$ transformations, we set $\eta(t,r)=0$ without loss of generality. The remaining gauge function $\xi^\mu(t,r)$ can then be properly chosen so that the PFV condition \eqref{fhvanish} can be satisfied. In fact, there is a class of admissible gauge choices, and within this class the resultant constraints on the series coefficients of the master functions are unique. Our choice is
\begin{align}
\Lambda_2(t,r){=}0,~M_0(t,r){=}0,~ M_2(t,r){=}0,~\eta(t,r){=}0,
\end{align}
\begin{align}
M_1(t,r)=\sum_{i=0,1}\mathrm{e}^{-\mathrm{i}\omega_{i}t}\Big[\frac{\beta^{(0)}_ip_{1\!-\!i}(\mu^2\!+\!2\!-\!2L^2\omega_i^2)}{(\mu^2+2)(p_{i}-p_{1\!-\!i})}\!+\!O\Big(\frac{1}{r^2}\Big)\Big]\,.
\end{align}
Under this additional gauge transformation, the PFV condition \eqref{fhvanish} can be implemented as long as the series coefficients in \eqref{PsiOddSeries}-\eqref{PsiEvenSeries} satisfy
\begin{align}
\alpha^{(0)}_i&=0\,,\quad i=0,1\,, \label{oddBC}
\\
\beta^{(1)}_i+\frac{p_{1-i}}{\mu^2}\beta^{(0)}_i&=0\,, \quad i=0,1,
\label{evenBC}
\end{align}
according to which, the metric and electromagnetic perturbations exhibit the following power law falloffs
\begin{align}
h^{\rm new}_{\mu\nu}&=
\begin{pmatrix}
	O(r^{-1}) & O(r^{-2}) & O(r^{-1}) &  O(r^{-1}) \\
	* & O(r^{-5}) & O(r^{-2}) & O(r^{-2}) \\
	* & * & O(r^{-1}) & 0 \\
	* & * & * & O(r^{-1}) \\
\end{pmatrix}\,,
\label{hPFV}
\\
\mathfrak{f}^{\rm new}_{\mu\nu}&=
\begin{pmatrix}
	\hspace{1.3em} 0\hspace{1.3em} & O(r^{-2}) & O(r^{-1}) &  O(r^{-1}) \\
	* & 0 & O(r^{-2}) & O(r^{-2}) \\
	* & * & 0 & O(r^{-1}) \\
	* & * & * & 0 \\
\end{pmatrix}\,,
\label{fPFV}
\end{align}
where the asterisks (*) denote components determined by their (anti-)symmetry properties. The decay profiles combine both odd and even parity contributions. Eq.~\eqref{oddBC} imposes a Dirichlet condition on each master function for odd-parity modes, while Eq.~\eqref{evenBC} imposes a Robin condition on each master function for even-parity modes. Notably, in the neutral limit $Q \to 0$, the Robin condition for the even-parity $i=1$ branch reduces to a Neumann condition, in which the derivative of the master function vanishes asymptotically. This completes the specification of the boundary conditions at infinity.

We now verify that the PFV condition indeed ensures the absence of electromagnetic energy flux at infinity, as mentioned in the introduction. To this end, consider the radial energy flux through a sphere of large $r$, given by
\begin{align}
\mathcal{F}|_{r}=\int_{S^2}\sqrt{-\tilde{g}}\,\tilde{T}^{r}_{\ t}d\theta d\phi\,.
\label{fluxdef}
\end{align}
Here, $\tilde{T}^{r}_{\ t}$ denotes the electromagnetic energy flux density in the radial direction, which includes contributions from both the background and perturbations. The dominant contributions to the flux come from products of electromagnetic perturbations. These include terms such as $r^2\mathfrak{f}_{t\theta}\mathfrak{f}_{r\theta}$ and $r^2\mathfrak{f}_{t\phi}\mathfrak{f}_{r\phi}$ in the integrand, leading to $\mathcal{F}|_{r}\sim O(r^{-1})$. Other contributions, including those involving metric perturbations and cross terms between background and perturbations, decay faster asymptotically. Consequently, the total flux $\mathcal{F}|_{r}$ vanishes in the limit $r\to\infty$, confirming that no residual electromagnetic flux leaks to infinity.

\subsection{Event horizon}

At the event horizon, we impose the ingoing wave condition to ensure that perturbations are purely infalling,
\begin{align}
\Psi^{\rm odd/even}_i(t,r) \sim \mathrm{e}^{-\mathrm{i}\,\omega_i (t+r_*)}\,,\, i=0,1\,,\quad r \to r_+\,. 
\end{align}
This selects the physically relevant solutions that guarantee causality. Together with the AdS boundary conditions, this completes the specification of boundary conditions for the master functions.

\section{Numerical method and results}
\label{Numerical-method-and-results}

To compute the quasinormal frequencies consistent with the boundary conditions derived in Sec.\ref{Boundary-conditions-on-master-functions}, we solve each master equation of \eqref{master-equations} numerically. Several numerical methods are available for asymptotically AdS spacetimes, including the Horowitz--Hubeny (HH) power-series method~\cite{Horowitz:1999jd}, the continued fraction method~\cite{Daghigh:2022uws}, the direct integration method~\cite{Wang:2017fie,Wang:2019qja}, and spectral method~\cite{Trefethen:2000,Fortuna:2022fdd}. In this work, we employ both the HH power-series method and a spectral method based on the Chebyshev discretization and differentiation to cross-check our results.

\begin{table*}[ht]
\caption{Gravitational QNMs for branch $i=0$ ($\omega_0$)\footnote{We note that the odd- and even-parity QNMs are not isospectral, but we do not discuss this here. Isospectral boundary conditions can be constructed via the Chandrasekhar--Darboux transformation, e.g., see Ref.~\cite{Moss:2001ga}.} and Electromagnetic QNMs for branch $i=1$ ($\omega_1$).}
\label{tb1QNM}
\begin{ruledtabular}
\begin{tabular}{ccccc}
$ n $ & $ \omega^{\rm odd}_{0}(r_+=1) $ & $ \omega^{\rm even}_{0}(r_+=1) $ & $ \omega^{\rm odd}_{1}(r_+=10) $ & $ \omega^{\rm even}_{1}(r_+=10) $ \\
\hline $ 0 $ & $ -2.0000\mathrm{i}$ & $ 2.1557-0.2855\mathrm{i} $ &  & $ -0.6340\mathrm{i} $ \\
$ 1 $ & $ 3.0331-2.4042\mathrm{i} $ & $ 3.4634-2.5734\mathrm{i} $ & $ -16.6233\mathrm{i} $ & $-13.8198\mathrm{i} $  \\
 &  &  & $ -25.8251\mathrm{i} $ &  \\
$ 2 $ & $ 4.9607-4.8982\mathrm{i} $ & $ 5.2304-4.9422\mathrm{i} $ & $ 12.4028-46.1091\mathrm{i} $ & $ 6.1567-34.1306\mathrm{i} $ \\
$ 3 $ & $ 6.9054-7.2897\mathrm{i} $ & $ 7.0965-7.3084\mathrm{i} $ & $ 24.7512-69.2446\mathrm{i} $ & $ 18.5372-57.7257\mathrm{i} $ \\
$ 4 $ & $ 8.8547-9.6604\mathrm{i} $ & $ 9.0022-9.6704\mathrm{i} $ & $ 37.3134-92.1505\mathrm{i} $ & $ 31.0146-80.7146\mathrm{i} $ \\
\end{tabular}
\end{ruledtabular}
\end{table*}

\begin{figure*}[htbp]
\includegraphics[totalheight=50mm]{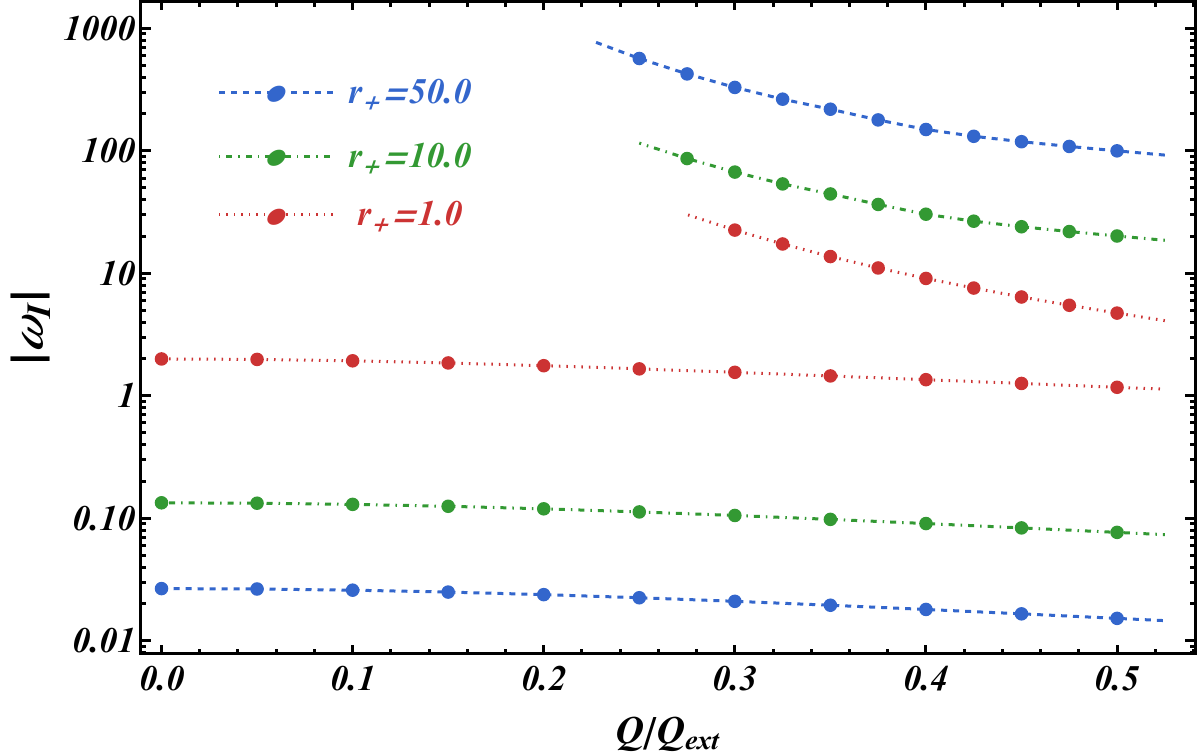}
\includegraphics[totalheight=51mm]{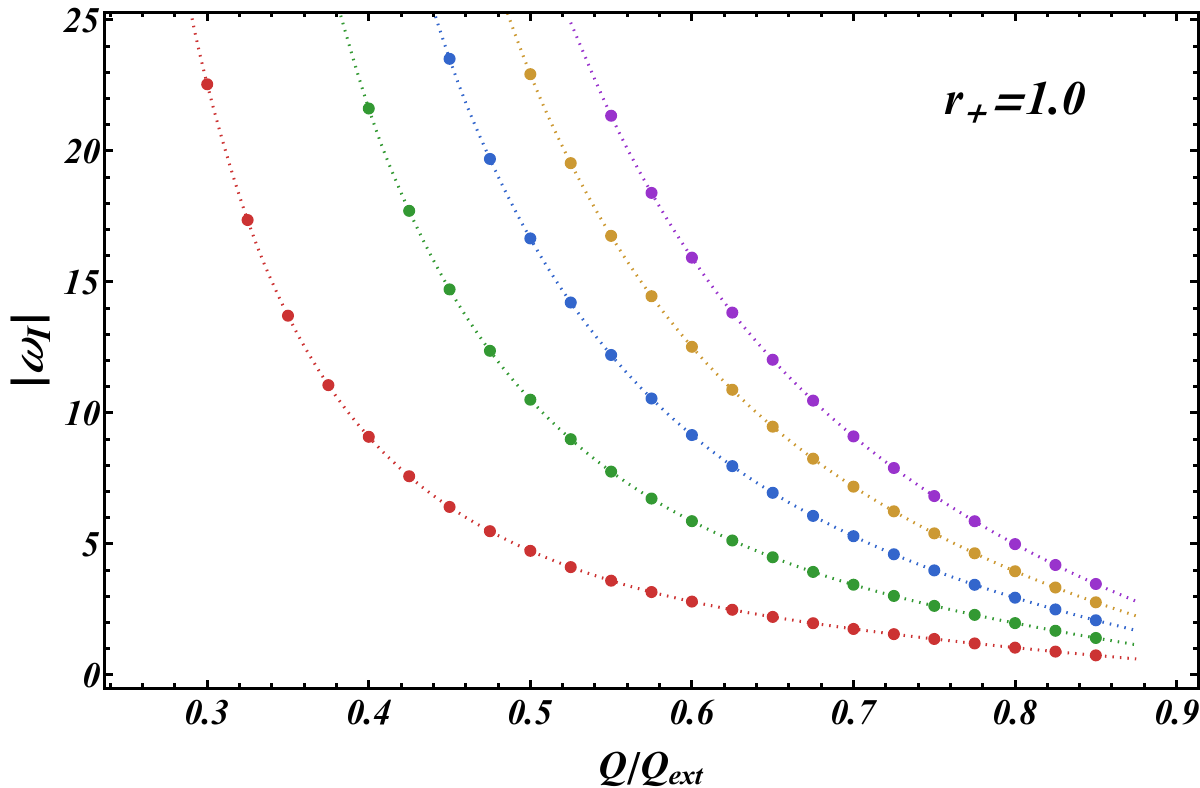}
\caption{\label{figPIqnf}
Purely imaginary-frequency branches of the master functions under variation of the normalized charge $Q/Q_{\text{ext}}$. Left: the two lowest purely imaginary-frequency branches of $\Psi^{\rm odd}_0$ for $r_+=1,10,50$. Right: purely imaginary-frequency branches of $\Psi^{\rm even}_0$ for $r_+=1.0$.
}
\end{figure*}

\subsection{Numerical implementation}

We begin with the spectral method implementation. This method~\cite{Trefethen:2000} provides high accuracy and exponential convergence for smooth eigenfunctions, allowing for precise determination of a large number of quasinormal frequencies. As the first step of this method, we factor out the near horizon behavior of a master function through the following procedure
\begin{align}
\Psi(t,r)=\mathrm{e}^{-\mathrm{i}\,\omega(t+r_*)}\phi(r)\,,\label{Psiexp}
\end{align}
where $\phi(r)$ is regular on the interval $(r_+,\infty)$. By introducing the compactified coordinate $u=r_+/r$, the domain $(r_+,\infty)$ maps to $u\in(0,1)$, with $u=1$ at the horizon and $u=0$ at the AdS boundary. A master equation for $\phi(u)$ then reduces to
\begin{align}
\mathcal{L}\,\phi(u)=0,\qquad u\in(0,1)\,,\label{eqLu}
\end{align}
where $\mathcal{L}$ is a second-order differential operator that depends linearly on $\omega$. We discretize $\phi(u)$ on a set of Chebyshev collocation points, approximating $\mathcal{L}$ by spectral differentiation matrices. This converts Eq.~\eqref{eqLu} into a generalized eigenvalue problem,
\begin{align}
\left[\mathcal{M}_0+\omega\,\mathcal{M}_1\right]\boldsymbol{\phi}=0\,,
\label{Mphi}
\end{align}
where $\mathcal{M}_0$ and $\mathcal{M}_1$ are matrices. The boundary conditions in Eqs.~\eqref{oddBC} and \eqref{evenBC} are originally formulated for the master function $\Psi(r)$. However, to apply our numerical scheme, we now need to translate them into algebraic constraints on the regular function $\phi(u)$ at $u=0$. These constraints have the form
\begin{align}
\text{odd}:&\quad \phi_i(u)=0,\quad i=0,1\,,
\\
\text{even}:&\quad \frac{d\phi_i(u)}{du}\!+\!\frac{1}{r_+}\big(\frac{p_{1\!-\!i}}{\mu^2}\!+\!\mathrm{i}\,\omega_i L^2\big)\,\phi_i(u)=0\,,\,i=0,1\,.
\end{align}
These relations are enforced by modifying the corresponding rows or columns of the matrices $\mathcal{M}_0$ and $\mathcal{M}_1$. Solving the resultant generalized eigenvalue problem yields a discrete set of complex frequencies $\omega$. Among them, the physically relevant quasinormal frequencies are identified as those with a negative imaginary part and exhibiting good numerical convergence and stability under the grid refinement.

For the HH power-series method, we employ an independent numerical implementation to compute the quasinormal frequencies. This method uses a Frobenius series expansion of the function $\phi(r)$, defined in Eq.~\eqref{Psiexp}, around the horizon $r=r_+$. A master function is expressed as
\begin{align}
\Psi(t,r)=\mathrm{e}^{-\mathrm{i}\,\omega(t+r_*)}\sum_{k=0}^{\infty}a_k(\omega)\Big(\frac{r-r_+}{r}\Big)^{k}\,,
\end{align}
where the coefficients $a_k(\omega)$ satisfy a recurrence relation derived from the master equation. Under this Frobenius series expansion, the boundary conditions in Eqs.~\eqref{oddBC} and \eqref{evenBC} translate into algebraic equations that determine the admissible quasinormal frequencies $\omega$. Explicitly, we obtain
\begin{align}
\text{odd}:&\quad \sum_{k=0}^{\infty}a_{k,i}(\omega_i)=0\,, \qquad i=0,1\,,
\\
\text{even}:&\quad \sum_{k=0}^{\infty}\Big(\frac{p_{1\!-\!i}}{\mu^2}\!+\!\mathrm{i}\,\omega_i L^2\!-\!r_+k\Big)\,a_{k,i}(\omega_i)=0\,, \quad i=0,1\,.
\end{align}
In practical computations, the infinite series is truncated at a sufficiently high order $k_{\mathrm{max}}$, and the roots $\omega$ are found by solving the resultant polynomial equation. The computation is repeated with increasing $k_{\mathrm{max}}$ until the values of $\omega$ converge with high numerical precision, ensuring the reliability of the identified quasinormal modes.

In all numerical computations, we fix the AdS radius to $L=1$ and focus on a representative angular index $\ell=2$ ($\mu^2=4$). Our results aim to show how the black hole parameters $r_+$ and $Q$ influence the quasinormal spectrum under the PFV conditions.

\subsection{Numerical results}

As a benchmark for subsequent investigation, we first compute the QNMs of uncharged black holes ($Q=0$). The results are listed in Table~\ref{tb1QNM} and can be directly compared with those reported in Ref.~\cite{Daghigh:2022uws,Michalogiorgakis:2006jc,Wang:2015goa}. For $\Psi^{\rm odd}_1$, the two purely imaginary frequencies $-16.6233\mathrm{i}$ and $-25.8251\mathrm{i}$ are both labeled as overtone $n=1$ due to a bifurcation phenomenon: as $r_+$ increases, the mode transitions from a complex frequency to a purely imaginary one and splits into two branches~\cite{Wang:2021upj}. This agreement serves as a nontrivial validation of our numerical programs and confirms that the PFV condition correctly reduces to the single-field perturbation boundary conditions in the uncharged limit.

\begin{figure*}
\centering
\includegraphics[width=0.48\textwidth]{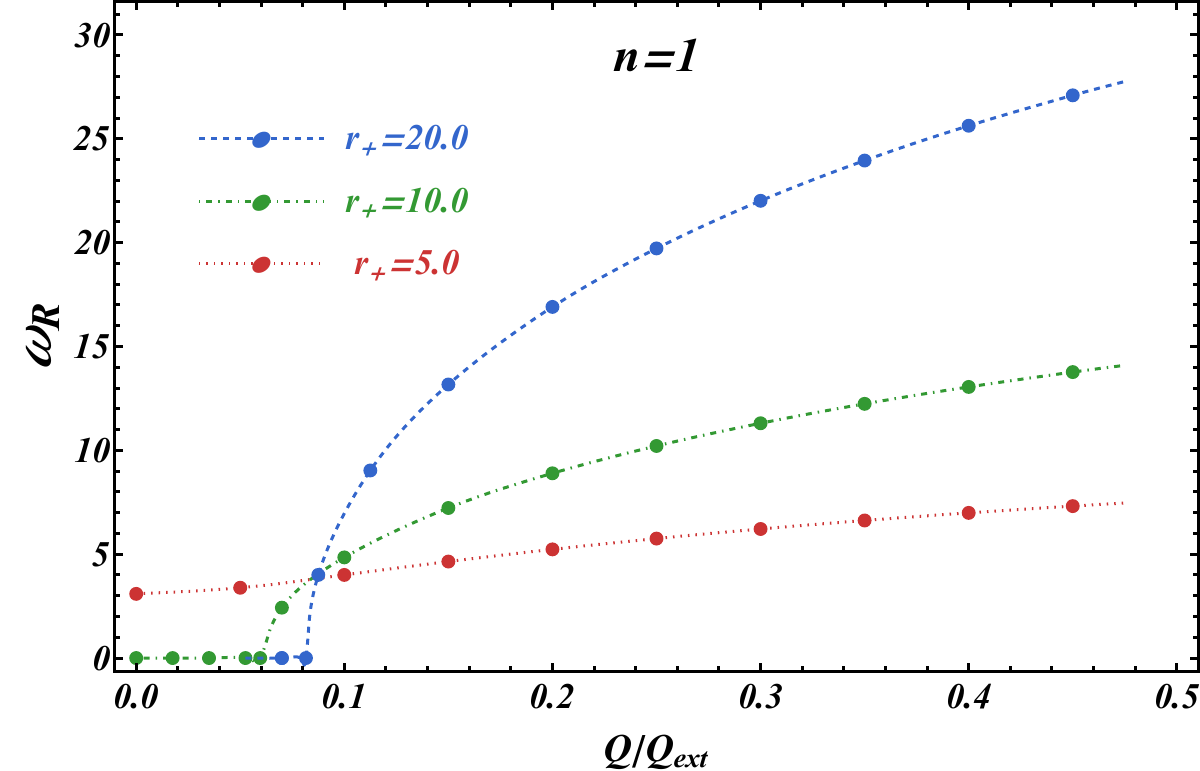}
\includegraphics[width=0.48\textwidth]{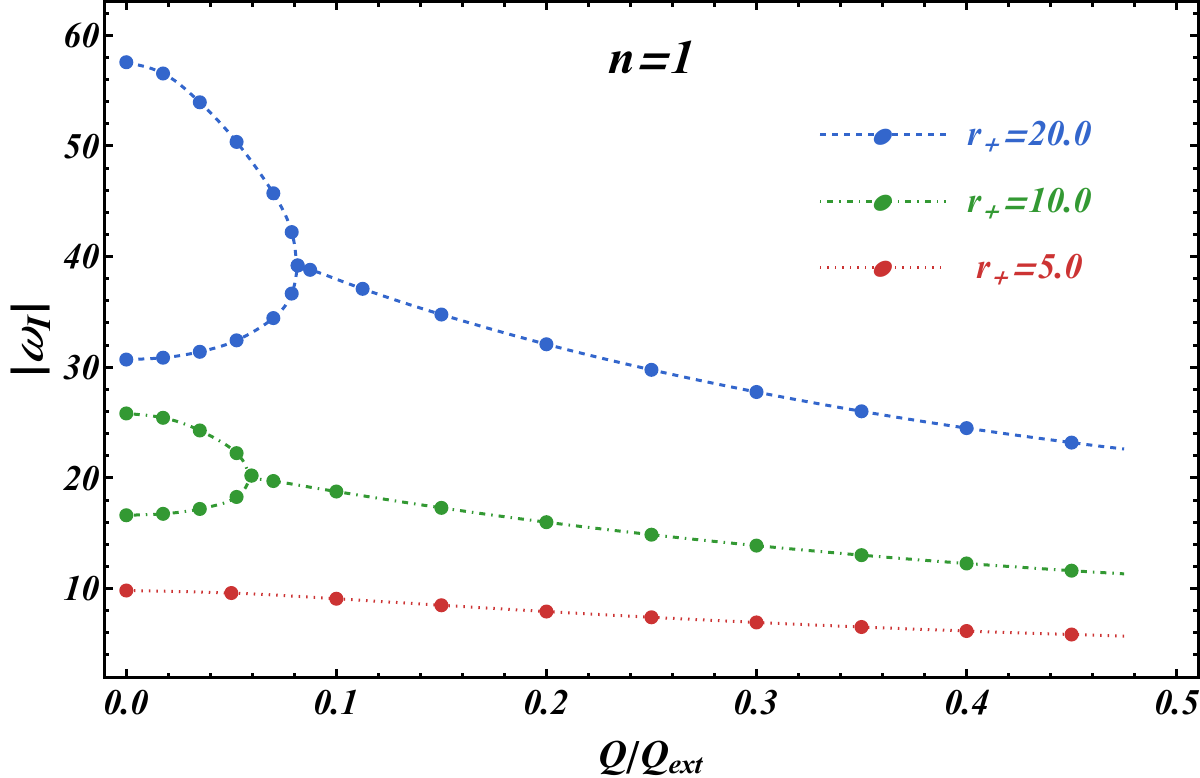}
\vspace{2mm}\\
\includegraphics[width=0.48\textwidth]{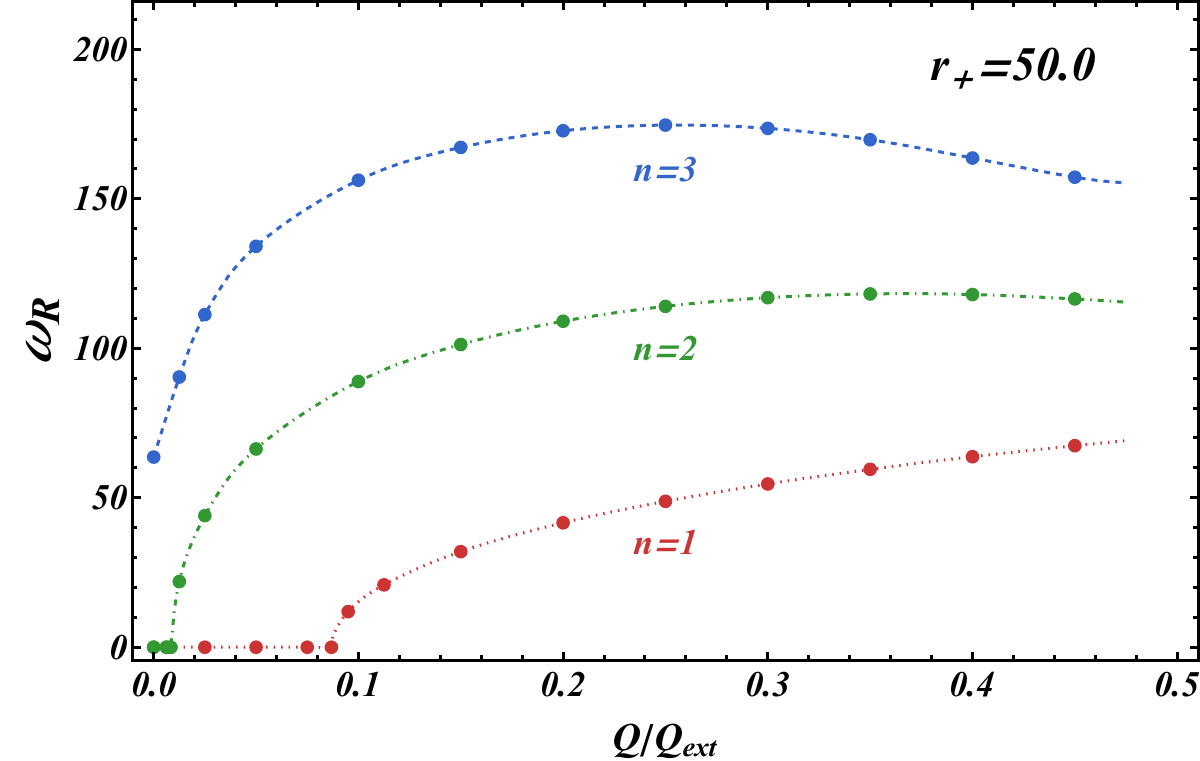}
\includegraphics[width=0.48\textwidth]{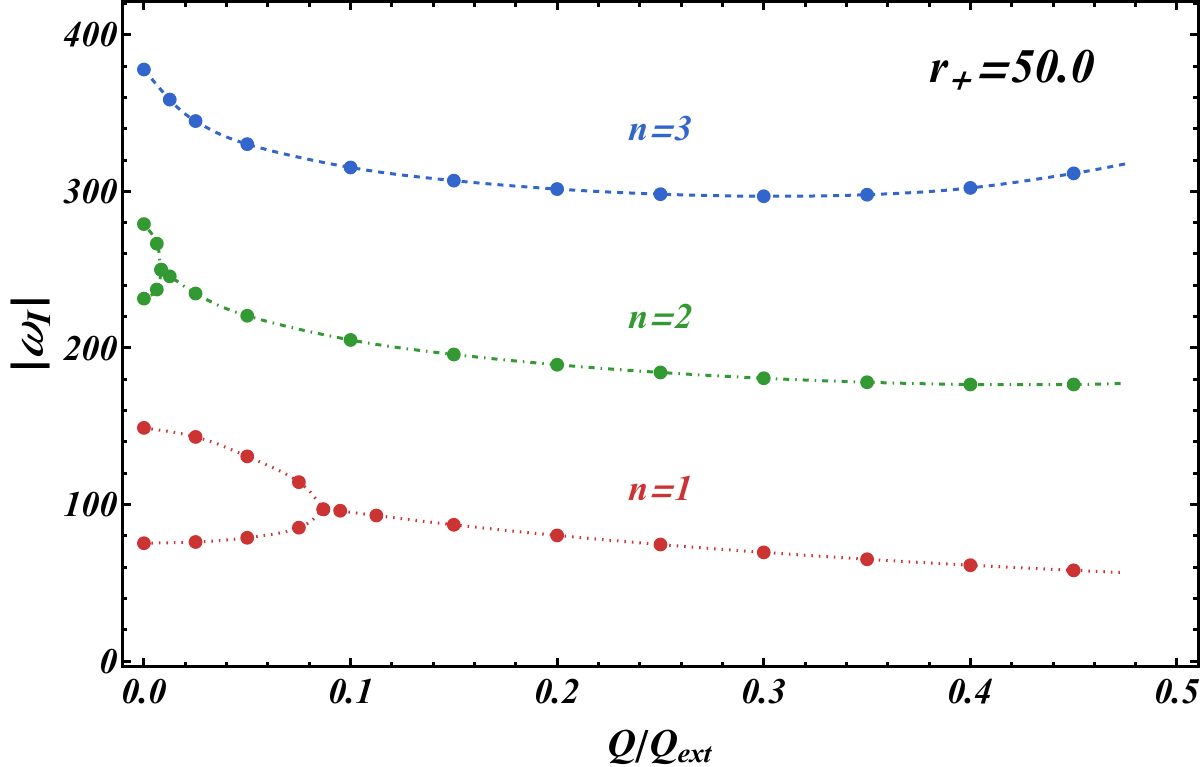}
\caption{\label{figodd1}
Quasinormal frequency branches of the master function $\Psi^{\rm odd}_1$ under variation of the normalized charge $Q/Q_{\text{ext}}$.
Upper panel: branches associated with the overtone $n=1$ for $r_+=5,10,20$.
Lower panel: branches associated with the overtones $n=1,2,3$ for $r_+=50$.
}
\end{figure*}

\begin{figure*}
\centering
\includegraphics[width=0.48\textwidth]{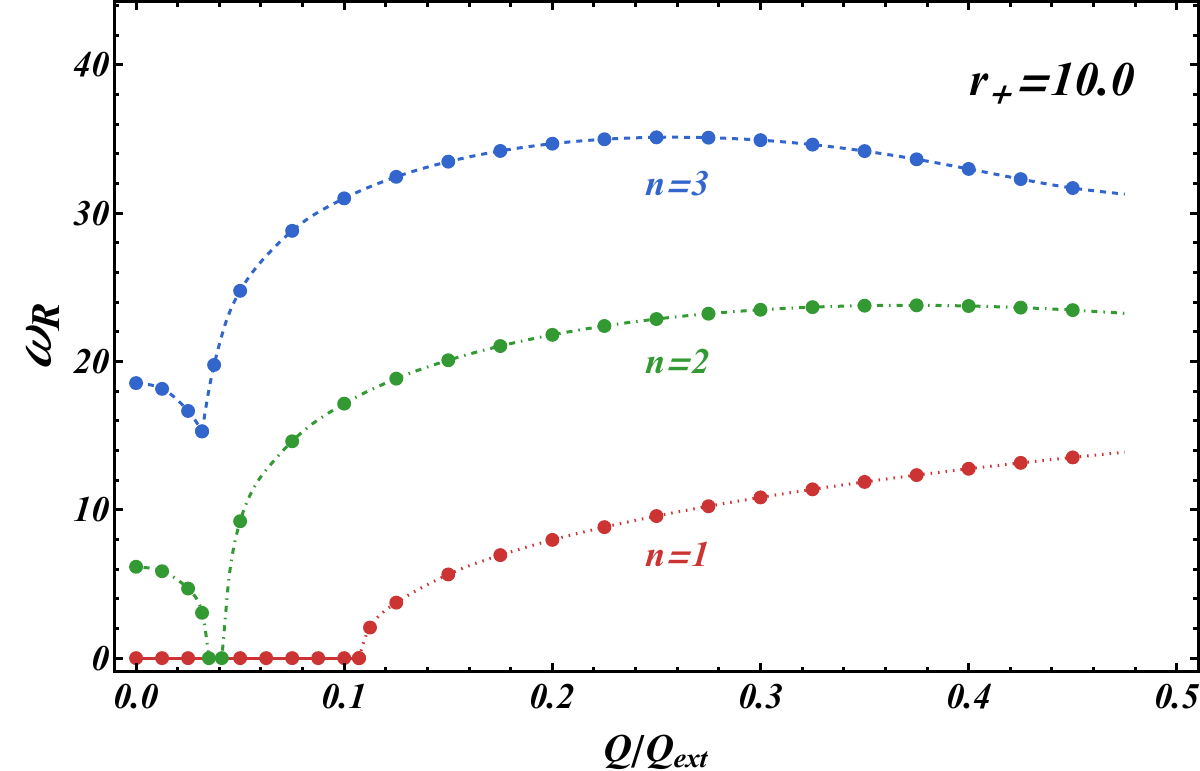}
\includegraphics[width=0.48\textwidth]{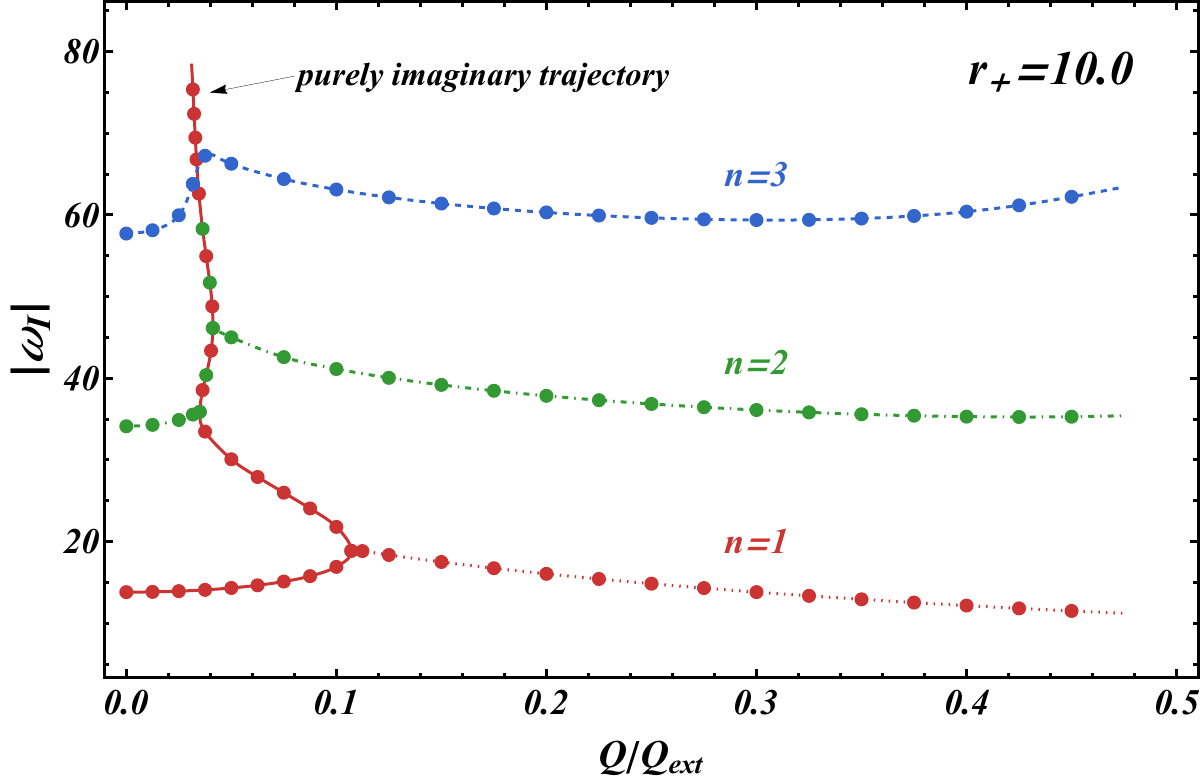}
\caption{\label{figeven1}
The real (left) and imaginary (right) parts of quasinormal frequency branches of the master function $\Psi^{\rm even}_1$
under variation of the normalized charge $Q/Q_{\text{ext}}$ for $r_+=10$.
The branches associated with the overtones $n=1$ and $n=2$ defined at $Q=0$ become connected
when both acquire purely imaginary frequencies, while the branch associated with $n=3$ remains isolated.
}
\end{figure*}

We now shift to the question of how black hole charge $Q$ influences the quasinormal spectrum. Throughout the following discussion, modes at finite $Q$ are identified by continuous deformation from the corresponding overtones defined at $Q=0$. Previous studies of RN--AdS QNMs typically imposed Dirichlet boundary conditions on the master functions, as in Ref.~\cite{Berti:2003ud}. In our prescription, the two parity sectors are associated with different types of boundary conditions, both derived from a common physical requirement. As a result, the odd-parity sector is governed by the same Dirichlet conditions employed in earlier analyses, whereas the even-parity sector is governed by Robin boundary conditions that have not been previously explored. Within this framework, our analysis of the charged spectrum reveals additional spectral features, including one that arises in the odd-parity sector, which has not been noticed or emphasized in previous studies.

When a nonzero charge is introduced, multiple purely imaginary modes emerge. Focusing on the $\Psi^\mathrm{odd}_0$ master function, the left panel of figure~\ref{figPIqnf} shows that, for each given horizon radius $r_+$, there are two lowest purely imaginary-frequency branches: one approaches the algebraically special frequency as $Q\to0$, the other diverges to large damping rate as $Q\to0$; since the figure focuses on the effects of varying $r_+$ on the quasinormal frequency, only one diverging branch per $r_+$ is plotted for clarity. Turning to the master function $\Psi^{\rm even}_0$, the right panel of figure~\ref{figPIqnf} illustrates multiple purely imaginary frequency branches that diverge as $Q\to0$, but there is no branch connecting to the algebraically special mode. This is because, for the even-parity perturbations, purely imaginary-frequency modes are absent in the uncharged limit and are replaced by low-lying modes~\cite{Michalogiorgakis:2006jc}. Together, the two panels of figure \ref{figPIqnf} highlight that the charges universally introduce numerous---possibly infinitely many---purely imaginary frequency branches that diverge as $Q\to0$, a characteristic observable in the spectrum of any master function.

Another important feature of the quasinormal spectrum is the charge-induced suppression of bifurcation. Figure~\ref{figodd1} shows the quasinormal frequencies of $\Psi^{\rm odd}_1$. The upper panel displays the behavior for different horizon radii $r_+$ at overtone $n=1$. As $r_+$ increases, the bifurcation becomes more pronounced, meaning that a larger normalized charge is required to suppress the bifurcation and merge the branches. The lower panel shows the frequencies for $r_+=50$ and overtones $n=1,2,3$. For higher overtone branches (e.g., $n=3$), the bifurcation is less noticeable, indicating that the suppression effect is more easily observed in lower overtone branches.

Additionally, we observed connectivities between spectral branches associated with different overtones defined at $Q=0$. Figure~\ref{figeven1} displays the quasinormal frequency obtained from $\Psi_1^\mathrm{even}$ by continuously deforming the $n=1,2,3$ modes, with horizon size $r_+=10$. Notably, the branches originating from $n=1$ and $n=2$ become connected when both modes acquire purely imaginary frequencies, forming a continuous trajectory that extends toward large damping. In contrast, the branch associated with $n=3$ remains separate and does not participate in this connection.

Overall, the introduction of charge qualitatively reshapes the quasinormal spectrum: it universally generates purely imaginary frequency branches, modifies bifurcation behavior in selected modes, and produces continuous trajectories linking branches associated with different uncharged overtones. These charge-induced features highlight the richer spectral structure of charged black holes.

\section{Discussion}
\label{Discussion}

In this work, we have introduced and implemented the PFV condition for linear perturbations of RN-AdS black holes. This condition requires both the metric and electromagnetic field-strength perturbations to vanish asymptotically at the AdS boundary, thus ensuring consistency with the geometric and physical structure of AdS spacetime. Using the reconstruction formulas, we translate the PFV condition into boundary conditions on the master functions: Dirichlet-type for odd-parity modes [see Eq.~\eqref{oddBC}] and Robin-type for even-parity modes [see Eq.~\eqref{evenBC}]. Our numerical analysis, employing the spectral and HH power-series methods, revealed charge-induced features in the quasinormal spectrum, such as the emergence of multiple purely imaginary modes and the modification or suppression of bifurcation behavior.

The key aspect of our approach is the reconstruction of perturbations, as shown in Eqs.~\eqref{h-recons}--\eqref{a-recons}. This reconstruction not only bridges the PFV condition and the boundary conditions on the master functions but also provides a foundation for future second-order perturbation studies. In such studies, quadratic combinations of the first-order perturbations will act as sources in the master equation, giving rise to nonlinear effects in charged AdS black holes. We also note that, unlike perturbations in vacuum black holes, the metric and electromagnetic perturbations exhibit nontrivial couplings between the two dynamical degrees of freedom, resulting in a mixture of QNMs from both. Recent spectral methods~\cite{Chung:2023zdq,Chung:2023wkd,Chung:2024ira,Chung:2024vaf,Chung:2025gyg} that extract QNMs directly from the metric perturbations could, in principle, be extended to such coupled systems. In this context, a proper identification and separation of the QNMs from different degrees of freedom would be a challenge.

While the spectral and HH power-series methods are effective for low-overtone modes, they exhibit reduced efficiency for modes with large imaginary parts as the overtone number increases. This limitation motivates the consideration of alternative methods, such as the continued fraction method~\cite{Leaver:1985ax,Nollert:1993zz}. This method has been proven highly effective for high-overtone modes under Dirichlet boundary conditions~\cite{Daghigh:2022uws}. However, its extension to Robin boundary conditions---relevant in this work---could enhance numerical stability and convergence for these QNMs.

In addition to addressing computational challenges, the PFV condition exhibits broader applicability. In particular, it may also be relevant for perturbations of higher-dimensional black holes in asymptotically AdS spacetimes~\cite{Konoplya:2003dd,Konoplya:2007jv,Konoplya:2008rq,Konoplya:2017zwo,Konoplya:2023kem}. It is expected to be relevant for perturbations in more general multifield systems, such as those in Bumblebee gravity~\cite{Liu:2023uft,Liu:2024wpa,Liu:2024axg,Deng:2025uvp,Li:2025itp} within asymptotically AdS backgrounds, where Lorentz symmetry breaking introduces additional couplings~\cite{Casana:2017jkc,Maluf:2020kgf}. In summary, this work clarifies the implementation of PFV conditions for linear perturbations of RN-AdS black holes and demonstrates its implications for calculations of QNMs. It also provides a basis for future studies on nonlinear perturbations and on extending the PFV condition to other perturbation systems in AdS spacetimes.

\begin{acknowledgments}

This work was supported by the Natural Science Foundation of China under Grant No. 11875082.

\end{acknowledgments}

%

\end{document}